\documentclass{aa}
\usepackage{epsfig}
\begin{document}

   \thesaurus{(11.06.1; 12.07.01)}
\title{A VLT colour image of the optical Einstein ring 0047--2808\thanks{Based
on observations obtained at the European Southern Observatory, Paranal, Chile, and the United Kingdom Infrared Telescope, Hawaii}}

\author{
S. J. Warren
\inst{1}
\and
G. F. Lewis
\inst{2}
\and
P. C. Hewett
\inst{3}
\and
P. M\o ller
\inst{4}
\and
P. Shaver
\inst{4}
\and
A. Iovino
\inst{5}
}

\offprints{S. J. Warren}

\institute{Blackett Laboratory, Imperial College of Science Technology and
Medicine, Prince Consort Rd, London SW7 2BZ, UK\\ 
(email: s.j.warren@ic.ac.uk)
\and Department of Physics and Astronomy, University of
Victoria, PO Box 3055, Victoria BC, V8W 3P7, Canada\\ 
(email: gfl@almuhit.phys.uvic.ca)\\ 
and Astronomy Department, University of Washington, Box 351580,
Seattle, WA 98195-1580 USA
\and Institute of Astronomy, Madingley Road, Cambridge CB3 0HA, UK
(email: phewett@ast.cam.ac.uk)
\and European Southern Observatory, Karl--Schwarzschild--Strasse 2,
D--85748 Garching bei M\"{u}nchen, Germany\\ 
(email: pmoller@eso.org, pshaver@eso.org)
\and Osservatorio Astronomico di Brera, Via Brera 28, I--20121 Milano, Italy \\
(email: iovino@brera.mi.astro.it)}
   \date{Received <date> / Accepted <date>}

   \maketitle

\begin{abstract}

The optical Einstein ring $0047-2808$ was imaged by the VLT UT1 during
the science verification programme. The ring is the image of a
high--redshift $z=3.595$ star--forming galaxy, with strong Ly$\alpha$
emission at $5589$\AA, gravitationally lensed by a massive early--type
galaxy at $z=0.485$. Relative to earlier NTT data the high
signal--to--noise ratio of the VLT Ly$\alpha$ narrow--band image allows
much improved constraints to be placed on the surface--brightness
profile of the source and on the mass, leading to a measured
mass--to--light ratio of ${\mathrm {\rm M}/{\rm L}_B \sim 13\,h}$ for the
deflector galaxy. We have combined the VLT $B$--band and Ly$\alpha$
narrow--band images with a $K$--band image obtained at UKIRT to produce
a deep colour image of the system.


\end{abstract}

\section{Introduction}

Einstein--ring gravitational lens images should be much more common at
optical wavelengths than at radio wavelengths (Miralda--Escud\'{e} and
Leh\'{a}r \cite{miralda}), but so far all but one of the known Einstein
rings have been discovered by radio techniques. The exception is
0047--2808 (Warren et al. \cite{warren_a}) where a high--redshift
$z=3.595$ star--forming galaxy, with strong Ly$\alpha$ emission at
$5589$\AA, is lensed by a massive early--type galaxy at $z=0.485$. We
are engaged in a survey to detect similar systems (Warren et al.
\cite{warren_b}). The search strategy is to identify anomalous emission
lines (Ly$\alpha$ from star--forming galaxies at $2<z<4$) in the
spectra of a large sample of distant early--type galaxies at $z\sim0.4$.
This has the advantage that by the very nature of the identification
procedure the redshifts of both the source and deflector are obtained
and so the full lensing geometry is known. In addition,
because the sources are extended the resulting images, rings or arcs,
offer the prospect of providing powerful constraints on the mass
distribution in the deflecting galaxies (Kochanek \cite{kochanek}).
Finally, because of the magnification, it is possible to study these
very faint sources both spectroscopically (Warren et al. \cite{warren})
and morphologically, resolving angular scales much smaller than is
possible for unlensed objects. The latter prospects are of particular
interest because the sources are similar to but fainter than the
population of high-redshift star--forming objects identified by Steidel
and coworkers (Steidel et al. \cite{steidel}), and cannot presently be
studied in any other way. In this paper we present VLT UT1 broad-- and
narrow--band imaging, together with UKIRT $K$--band imaging of
0047--2808.

\section{Observations and data reduction}

\begin{figure*}
\vspace{20 cm}
\caption[]{True-colour image of the field of the Einstein--ring
gravitational lens 0047--2808, showing a region
$68\arcsec\times81\arcsec$. North is up and East to the left. The image is a
{\it rgb} combination of the VLT $B$ and narrow-band images and the
UKIRT $K$ image. Because of the strong Ly$\alpha$ emission line in the
narrow-band the ring ($z=3.595$) appears as green. The lensing galaxy
($z=0.485$) is red because the galaxy is bright in the $K$ band but
faint in the $B$ and narrow bands. There is a second massive
elliptical galaxy in the field, visible as the brightest red image in
the SW quadrant, which has a redshift $z=0.488$.}
\end{figure*}

\subsection{VLT imaging}
Broad-band $B$ and Ly$\alpha$ narrow-band images of 0047--2808 were
obtained on the night of 1998 August 30, with the VLT UT1 test camera
as part of the Science Verification programme (Leibundgut et al.
\cite{leibundgut}). The CCD pixel size is $0\farcs045$. However the
CCD was binned $2\times 2$, so the pixel size in all the frames was 
$0\farcs09$. The narrow-band filter has a central wavelength $5589$\AA \
and width $20$\AA \ FWHM. Integration times were $3 \times300\,$sec ($B$)
and $3 \times 1200\,$sec (Ly$\alpha$). The seeing was $0.6-0.7\arcsec$.
Procedures followed for bias subtraction and flatfielding were
mostly standard. However the flatfielded narrow-band images required
a correction for large-scale gradients. This was achieved by firstly
combining the deregistered frames, clipping out objects. The
resulting frame was smoothed, and normalised, and the flatfielded
frames were divided by this correction frame.

\subsection{UKIRT imaging}

Broad-band $K$ images of 0047--2808 were obtained with the UKIRT
IRCAM3 instrument on the nights of 1997 September 12 and 13. The pixel
size was $0\farcs286$. The final image is a mosaic from two positions,
one centred on 0047--2808, total integration time $15\,120\,$sec, and
another at a position $58\arcsec$ to the SSW centred on a second
distant elliptical, total integration time $4725\,$sec. At each
position several sequences of 9--point dithers were summed.  The
seeing averaged $0\farcs7$. The data were flat--fielded using a
sequence of twilight sky exposures, and then an appropriate sky frame,
formed from a running median filter through the stack of images, was
subtracted from each data frame, and the resulting frames registered
and summed.

\subsection{Results}

Fig. 1 shows the {\it rgb} colour image resulting from combining the
$B$ (={\it b}), Ly$\alpha$ (={\it g}), and $K$ (={\it r}) images. The
ring stands out strongly in green because of the strong Ly$\alpha$
line in the narrow-band filter, while the lensing galaxy is very red
and is visible inside the ring. A minimum $\chi^2$ fit of a de
Vaucouleurs model for the light profile of the lensing galaxy in the
$K$--band image was computed by convolving two-dimensional $r^{1/4}$
profiles with the psf, measured from a star in the frame. (The
$K$--band image is best for fitting the galaxy profile because the
ring is not detected at this wavelength, and the contrast between the
galaxy and the ring is maximised.) The model was then convolved with
the Ly$\alpha$--band psf, scaled to the central counts in the
Ly$\alpha$ image, and subtracted. The resulting image of the ring,
rebinned to a pixel size of $0\farcs18$, is shown in the top left-hand
panel of Fig. 2.

\section{Gravitational lens model}

Compared with the original NTT image (Warren et al. \cite{warren_a})
the new VLT image has much higher signal--to--noise ratio. The counter
image predicted by our original model, but not convincingly detected
in the NTT image, is now clearly seen. The same modelling procedure as
described in Warren et al. (\cite{warren_a}), where the projected
surface mass density was assumed to follow the (intrinsic) de
Vaucouleurs profile, now measured from the $K$--band image (as
described above), has been applied. The single free parameter is the
global mass--to--light ratio (M/L).

\begin{figure*}
\vspace{20 cm}
\caption[]{False-colour images displaying the results of modelling the
mass distribution in the lens. The size of each square image is
$5.5\arcsec$ on a side. North is up and East to the left. The top LH image
shows the VLT Ly$\alpha$ image after subtraction of the image of the
lensing galaxy, and rebinned to a pixel size of $0\farcs18$. The lower
LH image shows the model of the source, and the 3--image and 5--image
caustics. The lower RH image shows the image that would be observed at
infinite spatial resolution, and the upper RH image is the lower RH
image convolved with the psf to emulate the observing conditions. The
upper RH image therefore models the upper LH image with zero
noise. The good correspondence of the two upper images gives a measure
of the accuracy of the mass model.}

\end{figure*}

Utilising  the   computational technique  for  arbitrary  lenses  with
elliptical symmetry   (Schramm~\cite{schramm})  the M/L ratio   in the
model was adjusted to produce  the most  compact configuration for the
unlensed image in  the source plane.  Here  we briefly  review the key
steps in the procedure; a position in the  image plane, represented by
the  complex coordinate  ${\rm z}$,  is mapped  onto the  source plane
position ${\rm \omega}$, according to

\begin{equation}
{\rm 
\omega(z,\bar{z}) = z - \nabla\Phi(z,\bar{z})\ ,
}
\label{mapping}
\end{equation}
where   ${\rm \nabla = 2\partial/\partial\bar{z}}$. The  deflection
potential,  ${\rm \Phi}$, is  related to  the  projected  surface mass
density in the lens, ${\rm \Sigma}$, by 
\begin{equation}
{\rm 
\nabla^2\Phi(z,\bar{z}) = 2\Sigma(z,\bar{z}) / \Sigma_{crit}
}
\label{gauss}
\end{equation}
where  ${\rm \Sigma_{crit}}$ is the critical surface mass density to
gravitational lensing and is given by
\begin{equation}
{\rm
\Sigma_{crit} = 
\frac{ c^2 }{4\pi G} \frac{D_{os}}{D_{ol} D_{ls}}}.
\label{critical}
\end{equation}
Here, ${\rm D_{ij}}$ are the angular diameter distances between the
source (s), lens (l) and observer (o) (Schneider et al. \cite{schneider}).

A 61$\times$61 pixel ($5.5\times5.5\arcsec$) region of the VLT
narrow--band image, centred on the galaxy, was used for the
computation. Assuming a fiducial value of M/L, the mapping given in
Equation~\ref{mapping} gives the coordinates in the source plane of
any image-plane coordinate. The M/L was adjusted, focusing the
emission over the source plane into a small region. This determines
the centroid of the source.  The source was then modelled as a
Gaussian profile. The structure in the ring (i.e. the angular extent
of the gaps, size of the counterimage) is dictated by the angular
extent of the source. A source of FWHM of $0\farcs2$ when reimaged by
the lensing potential was found to reproduce the structure in the ring
well. To reimage the source each pixel was sub--pixelated into a
10$\times$10 grid; these grid points were mapped to the source plane
to measure the surface brightness at each grid point.  Mapping of the
surface brightness in this way is accurate provided the grid spacing
mapped to the source plane is substantially smaller than the scale
over which the surface brightness of the source varies.

Having fixed the source position and profile the lens M/L was then finely
readjusted to provide the best fit, in terms of $\chi^2$, of the model
of the ring to the data. The results of this procedure are presented
in Fig. 2. The upper left--hand panel presents the VLT image of the
ring after subtraction of the model for the surface brightness
distribution of the foreground galaxy. Below, the model source is
shown on the same scale together with the caustic lines defined by the
gravitational lens model; the caustics delineate the regions of
multiple imaging over the source plane. The resultant image for this
source configuration is presented in the lower right--hand panel. In
the upper right--hand panel this image has been convolved with a
Gaussian seeing profile to approximate the observing conditions. There
is good correspondence between the structure in the model and the
observed structure of the ring.

The measured angular radius of the ring in the VLT image is
$1\farcs08\pm0.03$. This more accurate value is smaller than the value
measured from the NTT image ($1\farcs35\pm0.1$) by $20\%$ and this
significantly lowers the mass estimate. Part of the discrepancy
between the two measurements is due to the fact that the ring is
elliptical in shape and that the counterimage (invisible in the old
data) lies on the minor axis i.e. the old value for the radius was
measured along the major axis of the ellipse. The two measurements are
consistent therefore. The computed mass within the Einstein radius is
$1.73 (1.95)\times10^{11} h^{-1}{\rm M_{\odot}}$ for $q_o = 0.5
(0.1)~\footnote{$h$ is the Hubble constant in units of $100\,{\rm
km\,s^{-1}}{\rm Mpc^{-1}}$}$. The uncertainty in the mass estimate
within the Einstein radius is dominated by the uncertainty in the
radius rather than the form of the mass profile. Changing the radius
by $0\farcs06$ (i.e. $2\sigma$) changes the computed mass by $10\%$.
The M/L ratio for the model, corrected for luminosity evolution (Paper
I), is $ {\rm M}/{\rm L}_{B(0)}=14.2\, h (12.1\,h)$.

\begin{acknowledgements}
We are grateful to the ESO UT1 Science Verification team for providing
the optical data. GFL acknowledges support from the Pacific Institute
for Mathematical Sciences (1998-1999). The authors acknowledge the data
and analysis facilities provided by the Starlink Project which is run
by CCLRC on behalf of PPARC.

\end{acknowledgements}

\end{document}